\documentclass{article}

\usepackage{microtype}
\usepackage{graphicx}
\usepackage{subfigure}
\usepackage{booktabs} 
\usepackage{multirow}
\usepackage{soul}

\usepackage{hyperref}

\usepackage[accepted]{icml2023}

\usepackage{amsmath}
\usepackage{amssymb}
\usepackage{mathtools}
\usepackage{amsthm}
\usepackage{bbm}

\usepackage[capitalize,noabbrev]{cleveref}
\usepackage{enumitem}

\theoremstyle{plain}

\theoremstyle{definition}

\theoremstyle{remark}

\usepackage[textsize=tiny]{todonotes}

\newcommand{\METHOD}{Master-ASR }
\newcommand{\METHODNS}{Master-ASR}
\newcommand{\LAYER}{Artisan Layer }
\newcommand{\LAYERNS}{Artisan Layer}
\newcommand{\MODULE}{Specialist Score }
\newcommand{\MODULENS}{Specialist Score}
\usepackage{ulem}

\newcommand{\cw}[1]{{\color{black}#1}}

\icmltitlerunning{\METHOD: Achieving Multilingual Scalability and Low-Resource Adaptation in ASR with Modular Learning}

\begin{document}
\vspace{-5em}
\twocolumn[
\icmltitle{\METHODNS: Achieving Multilingual Scalability and Low-Resource \\Adaptation in ASR with Modular Learning}
 \vspace{-1em}



\begin{icmlauthorlist}
\icmlauthor{Zhongzhi Yu}{gt}
\icmlauthor{Yang Zhang}{ibm}
\icmlauthor{Kaizhi Qian}{ibm}
\icmlauthor{Yonggan Fu}{gt}
\icmlauthor{Yingyan (Celine) Lin}{gt}
\end{icmlauthorlist}

\icmlaffiliation{gt}{School of Computer Science, Georgia Institute of Technology, Atlanta, USA}
\icmlaffiliation{ibm}{MIT-IBM Watson AI Lab, Boston, USA}

\icmlcorrespondingauthor{Yingyan (Celine) Lin}{celine.lin@gatech.edu}

\icmlkeywords{Automatic Speech Recognition, Modular Model}

\vskip 0.2in
]



\printAffiliationsAndNotice{}  

\vspace{-0.5em}
\begin{abstract}
\vspace{-0.3em}
Despite the impressive performance recently achieved by automatic speech recognition (ASR), we observe two primary challenges that hinder its broader applications: (1) The difficulty of introducing scalability into the model to support more languages with limited training, inference, and storage overhead; (2) The low-resource adaptation ability that enables effective low-resource adaptation while avoiding over-fitting and catastrophic forgetting issues. Inspired by recent findings, we hypothesize that we can address the above challenges with modules widely shared across languages. To this end, we propose an ASR framework, dubbed \METHODNS, that, \textit{for the first time}, simultaneously achieves strong multilingual scalability and low-resource adaptation ability thanks to its modularize-then-assemble strategy. Specifically, \METHOD learns a small set of generalizable sub-modules and adaptively assembles them for different languages to reduce the multilingual overhead and enable effective knowledge transfer for low-resource adaptation. Extensive experiments and visualizations demonstrate that \METHOD can effectively discover language similarity and improve multilingual and low-resource ASR performance over state-of-the-art (SOTA) methods, e.g., under multilingual-ASR, our framework achieves a 0.13$\sim$2.41 lower character error rate (CER) with 30\% smaller inference overhead over SOTA solutions on multilingual ASR and a comparable CER, with nearly 50 times fewer trainable parameters over SOTA solutions on low-resource tuning, respectively. 
\end{abstract}
\vspace{-2.2em}
\section{Introduction}
Recent breakthroughs in deep neural networks (DNNs) have significantly advanced the performance of automatic speech recognition (ASR) in various applications under monolingual scenarios equipped with sufficient resources (i.e., sufficient labeled training data)~\cite{hsu2021hubert,baevski2020wav2vec,ao2021speecht5,babu2021xls,conneau2020unsupervised}. However, how to achieve comparable performance under more practical situations where there are fewer resources available, and multiple target languages need to be simultaneously supported, still remains an open question~\cite{babu2021xls,yadav2022survey}. Specifically, there are two critical challenges: 

\vspace{-0.5em}
\textbf{The multilingual scalability:} An ideal ASR system should be able to support multiple languages, \uline{while avoiding excessive overhead in terms of the training, inference, or model storage cost} when the number of supported languages increases~\cite{yadav2022survey}. To avoid the need for training completely different models for different languages~\cite{babu2021xls,conneau2020unsupervised}, the majority of existing works either introduce an adapter-like module to adapt the pretrained model to different languages with fewer additional model parameters~\cite{le2021lightweight,hou2021exploiting,fu2022losses}, or use a much larger model with a dedicated training recipe to increase the model capacity and cater to more complex multilingual ASR tasks~\cite{li2021scaling,li2022massively,pratap2020massively}. However, these approaches either require the model to be tuned for each language separately, resulting in high training costs~\cite{le2021lightweight,hou2021exploiting,fu2022losses}, or result in a significant increase in inference cost due to the larger model size~\cite{li2021scaling,li2022massively,pratap2020massively}. 

\vspace{-0.5em}
\textbf{The low-resource adaptation ability:} Given the limited training data from low-resource languages (e.g., less than one hour per language as in~\cite{fu2022losses}), effectively adapting the ASR model to target low-resource languages has been a long-lasting challenge in ASR. Existing attempts to address this challenge involve leveraging learned knowledge from pretrained models. In addition to directly tuning a pretrained model to low-resource languages~\cite{hsu2021hubert,baevski2020wav2vec,conneau2020unsupervised}, techniques such as utilizing more data from other modalities~\cite{zheng2021wav,Du2022ACJ,liang2020learning}, meta-learning~\cite{hsu2020meta}, and parameter-efficient tuning~\cite{fu2022losses,hou2021exploiting} are also used to further improve low-resource adaptation ability. However, how to better utilize the learned knowledge and avoid the issues of over-fitting~\cite{hou2021exploiting,cai2014stochastic} and catastrophic forgetting~\cite{winata2020meta,kessler2021continual} during adaptation remains an open research question.

\vspace{-0.4em}
Inspired by recent findings on the high similarity between ASR models trained for different languages~\cite{fu2022losses,lai2021parp}, we hypothesize that despite the differences between two languages, there is still sufficient similarity at a certain level (e.g., two languages may have significantly different phonemes but share similar morphemes), which can be leveraged to train generalizable sub-modules that can be shared across multiple languages. This approach has the potential to address both of the aforementioned challenges in ASR systems by sharing such sub-modules across different groups of languages at different layers. By adaptively combining different sub-modules, (1) the model's capacity can be improved to satisfy the need for a complex large-scale multi-lingual ASR system at limited training/inference/storage cost, and (2) the learned knowledge in such sub-modules can be effectively shared with new low-resource languages, avoiding the issues of over-fitting and catastrophic forgetting. 


\vspace{-0.4em}
Based on the above hypothesis and analysis, we make the following contributions in this paper: 
\vspace{-0.5em}
\begin{itemize}[leftmargin=*,topsep=0mm]
    \item We propose an ASR framework, dubbed \METHODNS, which addresses the aforementioned bottleneck challenges in multilingual ASR through a modularize-then-assemble approach. Specifically, \METHOD learns (1) a set of generalizable sub-modules, with each sub-module specializing in a different sub-task; (2) an assembly strategy that maps each supported language to the corresponding generalizable sub-modules in an end-to-end manner. 
    \vspace{-0.8em}
    \item We propose an efficient and effective low-resource adaptation approach in our \METHOD framework by only learning a new reassembly strategy for pretrained sub-modules without changing the sub-modules themselves. This approach avoids the catastrophic forgetting issue by preserving the pretrained sub-modules during adaptation and avoids the over-fitting issue by  reassembling the sub-modules, which enforces strong regularization. 
    \vspace{-0.8em}
    \item Extensive experiments and visualizations validate that \METHOD can effectively alleviate the aforementioned bottleneck challenge in ASR. In particular, \METHOD can discover language similarity and improve multilingual and low-resource ASR performance over state-of-the-art (SOTA) methods, e.g., a 0.13$\sim$2.41 lower character error rate (CER) with 30\% less inference overhead over SOTA solutions on multilingual ASR and a comparable CER with nearly 50 times fewer trainable parameters over SOTA solutions on low-resource tuning, respectively.  
\end{itemize}

\vspace{-0.8em}
\section{Related Works}
\vspace{-0.1em}
\subsection{Multilingual ASR}
\vspace{-0.3em}
Equipping ASR systems with the ability to deal with multilingual inputs without excessive training, inference, and storage overhead is a critical challenge in ASR~\cite{yadav2022survey,toshniwal2018multilingual,pratap2020massively,li2021scaling}. Existing multilingual ASR models mostly follow the pretraining-then-finetune pipeline~\cite{babu2021xls,baevski2020wav2vec,hsu2021hubert} where the model is first pretrained on a large multilingual dataset in a self-supervised manner and then tuned to target languages~\cite{conneau2020unsupervised,babu2021xls,hsu2021hubert}. However, the above pipeline is only effective when there are a limited number of languages. To support more languages, one natural way is to train a dedicated model for each language, which however will lead to training and storage costs that are proportional to the number of languages. As such, most of the existing methods either add a low-cost language-specific module to tune the model for each language~\cite{fu2022losses,hou2021exploiting,le2021lightweight}, or use a larger model with higher capacity and a dedicated training recipe to support more complex multilingual ASR tasks~\cite{li2021scaling,li2022massively,pratap2020massively}. However, these approaches either introduce additional training costs to tune the language-specific module on each target language or lead to increased inference costs due to the increased model size. Recent works try to use different modules for different languages; for example, \cite{nguyen2022refining} proposes to use language-specific fully-connected layers in the feed-forward network with a shared attention module to support multilingual processing, while \cite{pham2022adaptive} proposes to use the weight factorization method to generate a set of language-specific weights with a shared 1-rank base. Despite these efforts, 
existing works fall short of alleviating the aforementioned bottleneck challenge of multilingual scalability, motivating our exploration in this direction.


\vspace{-0.8em}
\subsection{Low-resource Adaptation in ASR}
\label{sec:low resource}
\vspace{-0.6em}
Exploring how to adapt an ASR model to a new language with limited labeled training data (low-resource language) is a long-standing challenge~\cite{Chen2015MultitaskLO,Deligne2001LowResourceSR,Miao2013DeepMN}. The key bottlenecking issues in low-resource adaptation are over-fitting~\cite{hou2021exploiting,cai2014stochastic} and catastrophic forgetting~\cite{winata2020meta,kessler2021continual}. Existing explorations can be categorized into three directions: (1) Constructing a better pretrained ASR model with more generalizable learned features and thus providing low-resource adaptation with a better starting point~\cite{baevski2020wav2vec,babu2021xls,ao2021speecht5,Du2022ACJ,zheng2021wav}; (2) Freezing the pretrained ASR model weights and introducing an additional module to adapt the ASR model to the target low-resource language, with commonly used modules including adapter tuning~\cite{hou2021exploiting,le2021lightweight,Cao2022AttentionFA} and mask tuning~\cite{fu2022losses,lai2021parp}; (3) Leveraging meta learning~\cite{Nichol2018OnFM,Finn2017ModelAgnosticMF} to generate a better initialization with a few data samples that have better adaptation ability~\cite{Kahn2019SelfTrainingFE,Li2019SemisupervisedTF}, with \cite{li2020towards,zhu2021multilingual} in particular, proposing to exploit the phoneme characteristics of different languages as prior knowledge to guide the prediction, providing a novel view in combining learning-based methods~\cite{graves2013hybrid} with statistical methods~\cite{ali1999acoustic}. However, these methods are still limited in handling both the catastrophic forgetting and over-fitting issues that arise during low-resource tuning of ASR. In particular, although freezing the pretrained ASR model weights and introducing an additional module~\cite{hou2021exploiting,le2021lightweight,Cao2022AttentionFA,fu2022losses} has the potential to overcome over-fitting and catastrophic forgetting issues by learning a combination of modules during tuning, its effectiveness in ASR is limited by the lack of an effective and efficient module design~\cite{hou2021exploiting,pham2022adaptive}.


\vspace{-0.8em}
\subsection{Modular Models}
\vspace{-0.6em}
Modular models learn a set of modules and a mapping strategy during training. This enables them to flexibly adopt 
 appropriate modules for different input data or target tasks~\cite{Kirsch2018ModularNL,Ponti2022CombiningMS,Crawshaw2020MultiTaskLW,Pan2020OnDA}. For example, \cite{Kirsch2018ModularNL} proposes a training method to effectively train a large model consisting of multiple modules and adaptively selecting different modules based on different given inputs and \cite{Ponti2022CombiningMS} proposes a novel model architecture to learn a set of LoRA adapters~\cite{hu2021lora} in a language model to simultaneously support multiple neural language processing tasks by adaptively selecting different combinations of LoRA adapters for different tasks. 
The merits of such models are two-fold: (1) They improve model capacity without increasing inference cost; (2) They help to decompose difficult tasks into simple sub-tasks, alleviating the learning difficulty and thus improving the achievable task accuracy~\cite{Kirsch2018ModularNL,Ponti2022CombiningMS}. Motivated by this, we hypothesize that such principles can be leveraged to improve both the multilingual scalability and low-resource adaptability of ASR systems. To the best of our knowledge, we are the first to explore the leveraging of the concept of modular models in designing scalable and data-efficient multilingual ASR models.

\vspace{-1.2em}
\section{Our Proposed \METHOD Framework}
\vspace{-0.3em}
\subsection{Problem Formulation}
\label{sec:prob}
\vspace{-0.6em}
We aim to develop a framework that can handle scalable multilingual ASR, while also supporting a data-efficient extension to low-resource languages. The problem we aim to solve can be described as follows: Given an initial training dataset $\mathcal{L}$ consisting of $L$ languages, we aim to develop an ASR model that can support all $L$ languages while also having the capability to support new low-resource languages without forgetting the previous $L$ languages. The latter means that given a new language $l'$, we can tune the model with only the data from $l'$ to enable the model simultaneously to support all the languages in the joint set $\mathcal{L}\cup l'$. 

\begin{figure}[t]
    \centering
    \includegraphics[width=\linewidth]{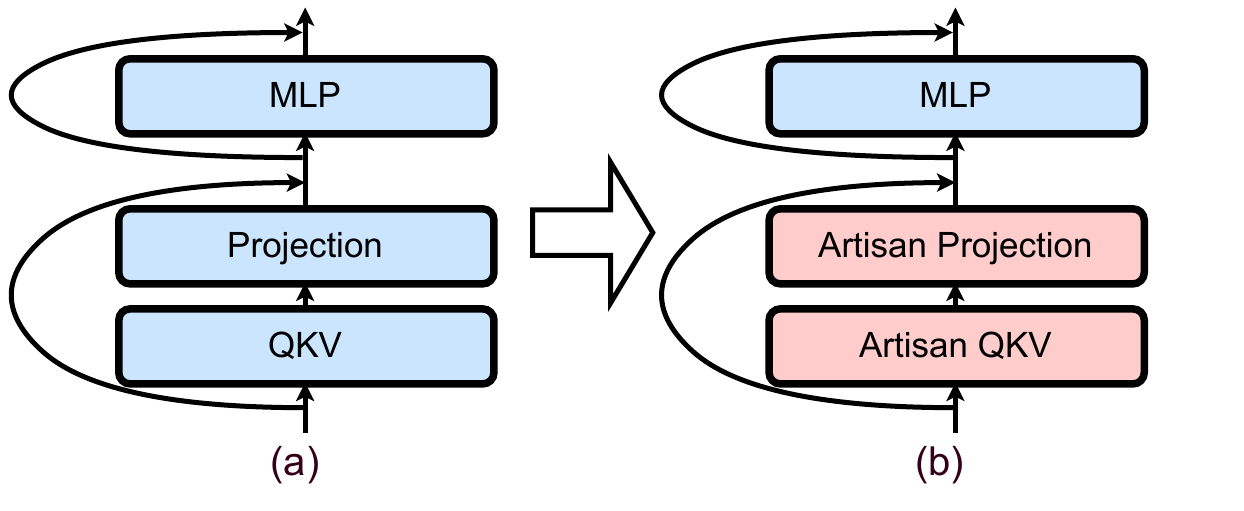}
    \vspace{-3em}
    \caption{An illustration of (a) a vanilla pretrained model; (b) a Master-ASR model built on top of the vanilla pretrained model by replacing the corresponding vanilla QKV or Projection layer with a new one, \LAYERNS. }
    \label{fig:overview}
    \vspace{-2.1em}
\end{figure}

\vspace{-0.7em}
\subsection{Drawn Inspirations from Previous Works}
\label{sec:motivate}
\vspace{-0.6em}
Recent advances in mask tuning techniques show that learning a set of masks on top of a self-supervised learning (SSL) pretrained ASR model can achieve promising recognition accuracy on monolingual ASR tasks~\cite{fu2022losses,lai2021parp}. Interestingly, these works find that masks learned for different languages share a high similarity. For example, learned masks for English, Spanish, and Russian all have more than 0.9 cosine similarity with each other~\cite{fu2022losses,lai2021parp}. This inspires us to think: Despite inherent differences among different languages, there is a potential to process different languages with highly similar modules. Although there exists such high similarity, existing methods still require independent training of different sets of masks for different languages, leading to non-trivial training and storage overhead in multilingual ASR scenarios and hindering low-resource ASR tuning from inheriting other languages' learned knowledge~\cite{fu2022losses,lai2021parp}. On the other hand, linguistic studies show that different languages share similarities at various levels. For example, Hebrew and Arabic share a high syntactic similarity but low phonetic similarity, while Spanish and Italian share a high phonetic similarity but low syntactic similarity~\cite{campbell2008ethnologue}. 

Motivated by the observations above, we aim to investigate if we can leverage such similarities (i.e., modular similarity and linguistic similarity) to adaptively share certain parts of an ASR model across languages. Specifically, our hypothesis is that there is a potential to first construct a set of generalizable sub-modules and then select a different combination of these sub-modules for different languages. 


\vspace{-0.7em}
\subsection{\METHODNS: Overview}
\begin{figure}
    \centering
    \vspace{1em}\includegraphics[width=0.85\linewidth]{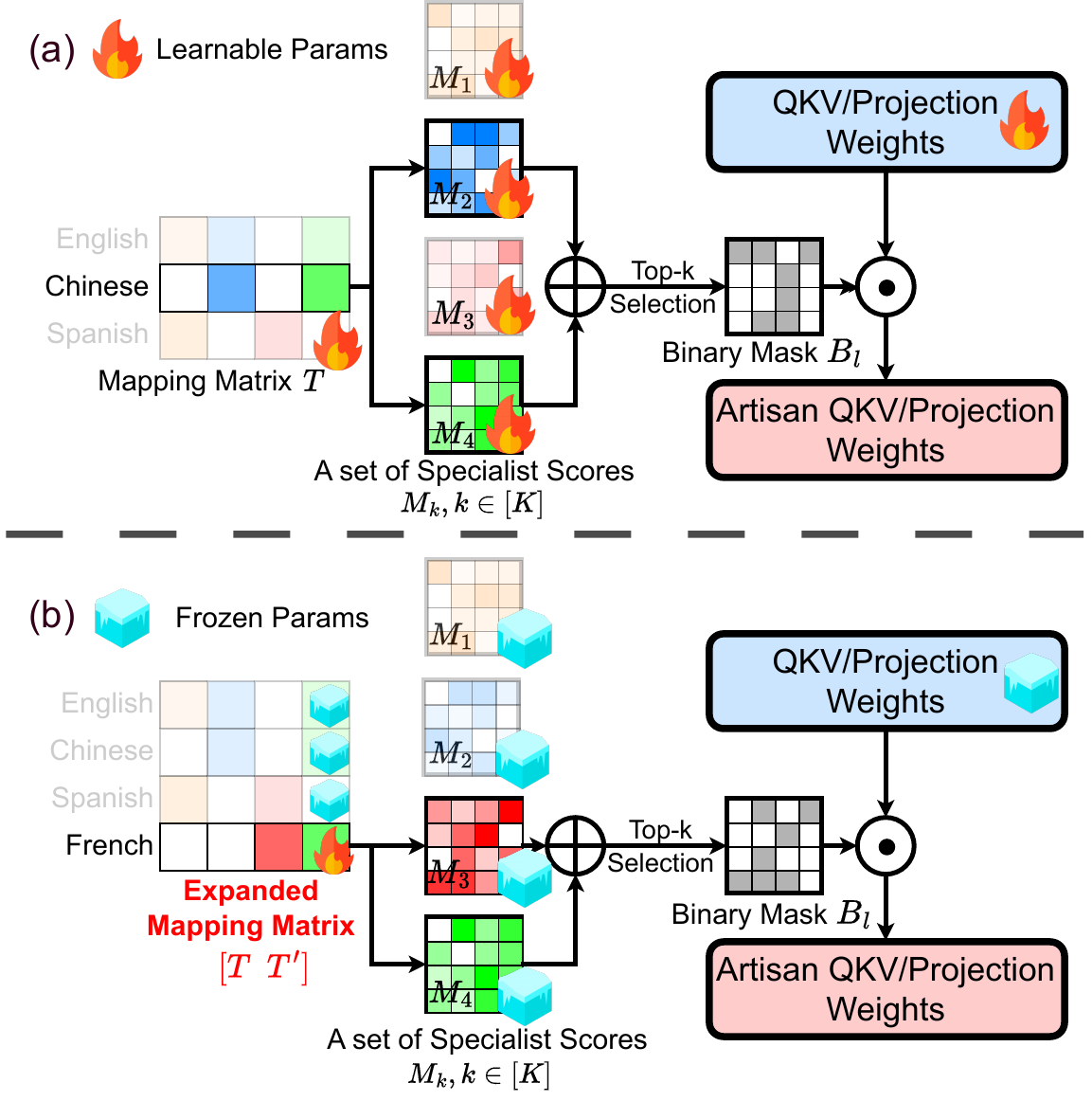}
    \vspace{-1.8em}
    \caption{Block diagram of the proposed \LAYER and our proposed two-stage training pipeline: \uline{\textbf{(a)}} Training Artisan Layer for scalable multilingual ASR, where we aim to learn (1) a mapping matrix $T$ and (2) a set of Specialist Scores $\{M_k\} (k\in[K])$, where $K=4$ in this example, and tune (3) the corresponding pretrained weights of the QKV or Projection layer; \uline{\textbf{(b)}} Tuning Artisan Layer for low-resource ASR, where we aim to support a new language by only inserting and tuning a new row in the mapping matrix while freezing all other parameters in the Artisan Layer.}
    \vspace{-1.8em}
    \label{fig:train_overview}
\end{figure}

\vspace{-0.3em}
Inspired by the aforementioned intriguing hypothesis, we develop our \METHOD framework that can adaptively share certain sub-modules across different languages. As shown in Fig.~\ref{fig:overview}, we replace the QKV and Projection layers in self-attention modules of a vanilla transformer with our proposed \LAYER (see Fig.~\ref{fig:train_overview} and Sec.~\ref{sec:model}). 
The purpose of the \LAYER is to learn shared weights across all the languages in the tuning dataset $\mathcal{L}$, while allowing different languages to select different sub-modules. Specifically, each \LAYER consists of three sets of parameters: (1) The pretrained weights inherited from the corresponding original QKV and Projection layer; (2) A set of \MODULENS s, each of which is of the same shape as the corresponding pretrained weights and can be adaptively combined to generate binary masks applied on top of the pretrained weights;
(3) A language-Specialist Score mapping matrix, of which the non-zero elements indicate the \MODULENS s (i.e., the corresponding mask scores) for a target language. 
Furthermore, to effectively train the above modules and matrices, Master-ASR integrates a two-stage training pipeline to (1) achieve multilingual ASR 
on dataset $\mathcal{L}$, i.e., the multilingual scalability (see Fig.~\ref{fig:train_overview} (a) and Sec.~\ref{sec:training}) and then (2) tune the trained multilingual ASR model on the newly added low-resource language $l'$, i.e., the low-resource adaptation ability. In this way, Master-ASR enables the trained model to extend the learned languages from multilingual dataset $\mathcal{L}$ to the joint set $\mathcal{L}\cup l'$ with minimal training, inference, and storage overhead (see Fig.~\ref{fig:train_overview} (b) and Sec.~\ref{sec:tuning}).

\vspace{-0.7em}
\subsection{\METHODNS: The \LAYER}
\label{sec:model}
\vspace{-0.3em}
In this subsection, we introduce the key building block, the \LAYERNS, in Master-ASR. As discussed in Sec.~\ref{sec:motivate}, we aim to design the \LAYER to fulfill two criteria: (1) It incorporates efficient sub-modules capable of adapting the outputs of the designed ASR model to different languages; (2) It can share these sub-modules adaptively across different languages based on their characteristics. In particular, the above two criteria are implemented on top of a vanilla QKV or Projection layer. As shown in Fig.~\ref{fig:train_overview} (a), the \LAYER first uses a mapping matrix $T$ to guide the adaptive summation of \MODULENS s to generate a distinct set of binary masks for different target languages. After that, \LAYER applies these generated binary masks to the pretrained weights of the corresponding QKV or Projection layer, adapting the model to different target languages. 

Formally, the design of the \LAYER can be described as follows: Given a QKV or Projection layer with a weight tensor $W\in\mathbb{R}^{c_{in}\times c_{out}}$, where $c_{in}$ and  $c_{out}$ are the number of input and output channels, respectively, the \LAYER introduces two additional components: (1) A set of $K$ \MODULENS s with each \MODULE $M_k\in\mathbb{R}^{c_{in}\times c_{out}} (k\in[K]$; $K$ is a hyperparameter in \METHODNS); (2) A mapping matrix $T\in\mathbb{R}^{L\times K}$, where the non-zero elements in $T$ indicate which \MODULENS s to use for the corresponding target language in $\mathcal{L}$. 
For a given language $l$, the \LAYER first generates the corresponding mask score $S_l$ by summing over a selected subset of the \MODULENS s, i.e.,
\vspace{-1.2em}
\begin{equation}
    S_l = \sum_{k=1}^K M_k \times \mathbbm{1}_{\sigma(T[l,k]) > 0.5},
\end{equation}
\vspace{-1.3em}

where $\mathbbm{1}_{f(.)}$ is an indicator function conditioned on $f(.)$ and $\sigma(.)$ is the Sigmoid function. Then for a target language $l$, given a preset sparsity ratio $t$ (e.g., $t=30\%$, which is a hyperparameter in \METHODNS), the corresponding binary mask $B_l\in \{0,1\}^{c_{in}\times c_{out}}$ is generated as follows. 

\vspace{-1.8em}
\begin{equation}
        B_l = \mathbbm{1}_{S_l > r},
\end{equation}
\vspace{-2em}

where $r$ is the $\lceil(1-t)\times c_{in}\times c_{out}\rceil$-th largest element in $S_l$ and $\lceil . \rceil$ is the ceiling operator. Finally, the weight tensor $W_{l}$ of the corresponding \LAYER is generated with 

\vspace{-1.8em}
\begin{equation}
    W_l = W \odot B_l,
\end{equation}

\vspace{-1.2em}
\noindent where $\odot$ is the element-wise product operator.

\vspace{-0.3em}
\subsection{\METHODNS: Training Towards Scalable Multilingual ASR}
\label{sec:training}
\vspace{-0.3em}
In the multilingual ASR training stage, we aim to train the \METHOD model to simultaneously achieve decent accuracy for all the $L$ languages. Our training objective can be described as follows: 

\vspace{-1.6em}
\begin{equation}
    \min_{\mathcal{W},\mathcal{T},\mathcal{M}} \sum_{l\cw{\in\mathcal{L}}}\sum_{(x,y)\in\mathcal{D}_l} \mathcal{J}\left(f(x;\mathcal{W},\mathcal{T},\mathcal{M}), y\right),
    \label{eq:train}
\end{equation}
\vspace{-1.6em}

where $\mathcal{J}(.)$ is the Connectionist Temporal Classification (CTC) loss, $(x, y)$ are the audio inputs and corresponding transcriptions of training dataset $\mathcal{D}_l$ corresponding to language $l$, and $f(.)$ is the \METHOD model parameterized by its three sets of parameters (1) $\mathcal{W}$ (the total set of vanilla Transformer weights), (2) $\mathcal{T}$ (the total set of mapping matrices), and (3) $\mathcal{M}$ (the total set of Specialist Scores).

While the above objective in Eq.~\ref{eq:train} can be optimized in an end-to-end manner,
effectively training the \LAYER towards its maximum potential is still a non-trivial task. In particular, there are two challenges: 
(1) \textbf{Collapse of $\mathcal{T}$}: A recent work shows that training a modular model with a mapping matrix can be problematic, as certain $T\in\mathcal{T}$ may collapse into a high entropy or non-sparse distribution~\cite{Ponti2022CombiningMS}. This issue hinders the model from learning distinct features across different modules (e.g., \MODULENS s in Fig.~\ref{fig:train_overview}), and thus, its capability to generate sufficiently different outputs for different languages; (2) \textbf{Mask convergence}: Recent works indicate that mask tuning requires a low-noise condition~\cite{lai2021parp,fu2022losses}, thus making it difficult to learn an optimal set of masks when the mapping matrix $T$ undergoes rapid changes during training. To tackle the two challenges above, \METHOD integrates the following techniques.

To tackle (1) \textbf{collapse of $\mathcal{T}$}, we manipulate the learning rate and the update frequency of all elements in $\mathcal{T}$. Specifically, we increase the learning rate of all $T\in\mathcal{T}$ to be larger than all other parameters in \METHOD (see Fig.~\ref{fig:train_overview}) by $\alpha$ times, and only update $T\in\mathcal{T}$ every $\beta$ iterations while all the other parameters are updated in each iteration. With a higher learning rate for $T$, we aim to  facilitate decisive selection of \MODULENS s during training, e.g., given a \MODULE $M_k$ and a target language $l$, $\sigma(T[l,k])\approx 0$ or $\sigma(T[l,k])\approx 1$. We empirically observe that doing so can avoid \METHOD from frequently alternating between selecting and deselecting a specific \MODULE for a given language in consecutive updates, as shown in Table~\ref{tab:alpha}. Such frequent switching could prevent the corresponding \MODULE from effectively learning a language-specific representation. On the other hand, the lower update frequency for $T$ can enable the selected \MODULENS s to undergo several updates before updating $T$. Our observation is that it can increase the standard deviation of $T$, suggesting $T$ can better determine the optimal selection of \MODULENS s for each language, as shown in Table~\ref{tab:alpha}.

To tackle (2) \textbf{mask convergence}, a widely adopted strategy that can alleviate the mask convergence issue is to adopt a prune-then-grow pipeline~\cite{lai2021parp,jaiswal2022training,liu2022don}. This pipeline first prunes less important weight elements by setting them to zero and then tunes the full model including these zeroed-out weight elements, providing them with a chance to grow back~\cite{lai2021parp,jaiswal2022training,liu2022don}. However, it is not straightforward to adopt this pipeline in \METHOD because its \METHOD model generates a distinct set of binary masks for each target language, i.e., having a total of $L$ sets of binary masks. Directly setting the less important weight elements to zero based on one set of binary masks will sabotage other sets of binary masks. To address this, we propose an iterative update strategy, where we alternate between updating $\mathcal{W}$ and $\mathcal{M}$ every $\gamma$ iterations. This strategy can not only train $\mathcal{M}$ to produce effective binary masks for different languages, but also adjust $\mathcal{W}$ to better accommodate the binary masks generated by $\mathcal{M}$, as shown in Table~\ref{tab:mask-convergence}.

\vspace{-1.0em}
\subsection{\METHODNS: Tuning on Low-resource Languages}
\label{sec:tuning}
\vspace{-0.5em}
In this section, we elaborate on how to leverage the trained multilingual Master-ASR model above in a low-resource tuning scenario.
As mentioned in Sec.~\ref{sec:low resource}, existing works need to either tune a full model~\cite{baevski2020wav2vec,babu2021xls} or train language-specific modules from scratch~\cite{fu2022losses,hou2021exploiting} to support a new low-resource language $l'$. Both have been shown to easily suffer from over-fitting and catastrophic forgetting issues~\cite{hou2021exploiting,cai2014stochastic,winata2020meta,kessler2021continual}. In contrast, in \METHODNS, the learned \MODULENS s in each \LAYER (see Sec.~\ref{sec:training}) provide a novel design knob to support $l'$. Specifically, this is achieved by learning a new combination of \MODULENS s in each \LAYERNS, i.e., inserting and optimizing an additional row in each mapping matrix $T\in\mathcal{T}$. 

Formally, during tuning, we first freeze all the parameters in the trained \METHOD model to preserve the knowledge learned from all the $L$ languages and then add two additional sets of parameters to the \METHOD model: (1) A randomly initialized classification layer ${W'}_{cls}$ for better adapting to the characteristics of language $l'$ and (2) an additional row $T'$ to be added to each of the mapping matrices, i.e., extending $T\in\mathcal{T}$ to $[T\ T']\in\mathcal{T}'$, where $T'$ represents the learnable \MODULENS s combination strategy for $l'$ and $\mathcal{T}'$ is the total set of extended mapping matrices. Given the training dataset $\mathcal{D}'$ corresponding to language $l'$, we aim to optimize the following object in this stage, 

\vspace{-2em}
\begin{equation}
     \min_{\mathcal{T}',{W'}_{cls}} \sum_{(x, y)\in\mathcal{D}'} \mathcal{J}(f(x;\mathcal{W}\cup\{{W'}_{cls}\}, \mathcal{T}', \mathcal{M}), y).
    \label{eq:tune}
\end{equation}
\vspace{-1.6em}

Thanks to our \LAYER design, the low resource tuning in \METHOD (i.e., the optimization of Eq.~\ref{eq:tune}) can be differentiably updated in an end-to-end manner. 

\begin{table*}[!htp]
\vspace{-0.6em}
    \caption{Benchmarking our \METHOD with SOTA multilingual ASR solutions. The accuracy of each language is measured in terms of CER, and the reported inference, training, and storage overhead are normalized to separate weight tuning. Column "All-avg" is the average CER achieved over 51 languages in our multilingual dataset. It is worth noting that all methods, except the Separate Weight Tuning baseline, in this table adopt a shared multilingual model to process all languages.}
    \vspace{0.2em}
    \centering
    \resizebox{\linewidth}{!}{
    \begin{tabular}{c|ccc|cccccccccc|c}
        \toprule
            Method & Inference & Training & Storage &de &zh &es &tt &ru &it &ky &tr &sv-SE &fr & All-avg \\
            \midrule
            Separate Weight Tuning & 1x & 1x & 1x & 14.45 & 21.53 & 9.59 & 10.11 & 14.77 & 9.99 & 14.63 & 10.33 & 18.22 & 22.63 & 12.71 \\
            Shared Weight Tuning & 1x & 0.57x & 0.02x & 23.43 & 31.15 & 14.33 & 16.29 & 22.47 & 14.75 & 20.64 & 16.39 & 27.81 & 32.10 & 19.48 \\
            \cite{fu2022losses} & 0.8x & 0.57x & 0.05x & 25.12 & 34.38 & 16.01 & 18.64 & 23.90 & 15.31 & 22.2 & 18.52 & 29.06 & 36.73 & 22.10 \\
            \cite{pham2022adaptive} & 1x & 0.57x & 0.02x & 18.51 & 25.39 & 12.06 & 12.21 & 19.84 & 12.64 & 19.04 & 13.23 & 23.56 & 29.37 & 16.65 \\
            \cite{nguyen2022refining} & 1x & 0.57x & 0.67x & 16.77 & 23.94 & 10.35 & 10.71 & 16.59 & 11.31 & 15.54 & 11.48 & 20.86 & 25.81 & 14.37 \\
            \METHODNS-Adapter & 1.02x & 0.57x & 0.04x & 18.27 & 25.09 & 11.88 & 12.04 & 19.78 & 12.56 & 18.36 & 13.19 & 23.67 & 28.59 & 17.35 \\\cmidrule{1-15}
            \METHOD & 0.7x & 0.57x & 0.04x & 16.31 & 23.53 & 10.14 & 10.67 & 15.98 & 10.84 & 15.49 & 11.08 & 20.54 & 25.52 & 14.24 \\
        \bottomrule
    \end{tabular}
    }
    \vspace{-1.5em}
    \label{tab:multilingual}
\end{table*}
\vspace{-1em}
\section{Experiments}
\vspace{-0.3em}
\subsection{Experiment Settings}
\vspace{-0.6em}
\textbf{Datasets and models}:
\uline{Datasets}. We evaluate \METHOD using a subset of the widely used large-scale CommonVoice dataset~\cite{ardila2019common}. Specifically, this subset \cw{comprises} 51 languages, each of which contains one hour of training data and one hour of validation data, to train our multilingual ASR model as described in Sec.~\ref{sec:training}. 
\cw{Furthermore, we collect an additional dataset consisting of six languages, with 10 minutes of training data and 10 minutes of validation data for each language, to evaluate the performance of low-resource tuning as discussed in Sec.\ref{sec:tuning}.}  
\uline{Models}. We implement \METHOD and baseline methods on a pretrained XLSR-53~\cite{conneau2020unsupervised} model without using a language model, such as a 4-gram language model~\cite{heafield2013scalable}, \cw{to ensure} a fair comparison. XLSR-53 is pretrained on 53 languages in an SSL manner. It is worth noting that the six-language low-resource tuning dataset we collected \cw{does not overlap with the dataset used} for XLSR-53 pretraining or the collected 51-language multilingual dataset.

\textbf{Multilingual ASR training settings:}
We design our training recipe \cw{following} the training schedule used in~\citet{baevski2020wav2vec}. Specifically, we train models for 100k iterations on 36 GPUs using an Adam optimizer with an initial learning rate of 5e-5 \cw{and} a tri-stage schedule for all modules except $\mathcal{T}$. Unless \cw{stated }otherwise, we set $t=0.3$, $\alpha=10$, $\beta=5$, and $\gamma=5,000$. 

\textbf{Low-resource fine-tuning settings:}
We follow the settings in~\citet{conneau2020unsupervised} to tune models on low-resource languages. Specifically, we tune models for 20k iterations with an Adam optimizer and an initial learning rate of 5e-5 plus a tri-stage schedule~\cite{baevski2020wav2vec}. During \cw{the} tuning \cw{process}, we first freeze the mapping matrix $\mathcal{T}$ and only tune the classification layer $\mathcal{W'}_{cls}$ for 2k iterations. \cw{Subsequently}, we \cw{incorporate} $\mathcal{T}$ into \cw{the} tuning \cw{process} and update it along with $\mathcal{W'}_{cls}$ and biases in each layer. We set the learning rate of $\mathcal{T}$ to be the same as other modules. 

\textbf{Baselines and evaluation metrics:}
Our baselines include SOTA ASR solutions and different competitive variants of our proposed method. Specifically, \uline{for multilingual ASR}, we consider a total of six baselines: (1) Two vanilla weight tuning methods (i.e., directly updating weight elements in a differentiable manner) on top of a weight-shared model and separate models across different languages, respectively; (2) Two SOTA modular-based multilingual ASR systems~\citet{pham2022adaptive} and~\citet{nguyen2022refining}; (3) Two variants of our proposed method, i.e., a variant that learns only one mask for each layer, as in~\citet{fu2022losses}, and another variant that uses the SOTA speech adapter~\cite{le2021lightweight} in the \LAYER instead of our proposed \MODULENS, dubbed \METHODNS-Adapter. \uline{For low-resource tuning}, we consider three baselines: vanilla weight tuning, mask tuning~\cite{fu2022losses}, and tuning with the SOTA speech adapter~\cite{le2021lightweight}. \uline{For evaluation metrics}, we comprehensively evaluate our method from different aspects: (1) The achievable accuracy measured with character-error-rate (CER); (2) The training cost calculated as the product of the number of GPUs and the number of iterations; (3) The inference cost  in terms of the required floating-point operations (FLOPs); (4) The storage cost quantified by the space needed to store the multilingual ASR model that supports $L$ languages; (5) The total number of trainable parameters.

\vspace{-0.8em}
\subsection{Benchmark \METHOD with SOTA ASR Systems}
\vspace{-0.3em}
\textbf{Benchmark on multilingual ASR.} We begin by evaluating the capability of \METHOD in multilingual ASR, comparing it with SOTA systems~\cite{pham2022adaptive,nguyen2022refining,fu2022losses} and a variant of \METHOD, namely the aforementioned \METHODNS -Adapter. As shown in Table~\ref{tab:multilingual}, \METHOD consistently outperforms all multilingual ASR baselines, achieving higher recognition accuracies with a CER reduction ranging from 0.13 to 7.86.
Moreover, although separate weight tuning yields the lowest CER, its training and storage costs are excessively high. This necessitates multilingual ASR systems, among which  Master-ASR achieves a triple-win in terms of recognition accuracy, training/inference speed, and required storage. 

\begin{table}[!tp]
\vspace{-0.8em}
    \caption{Benchmarking our \METHOD on low-resource tuning with SOTA solutions. Each language is trained with only 10-min data. ``Param." indicates the number of trainable parameters.}
    \centering
    \resizebox{\linewidth}{!}{
    \begin{tabular}{c|c|cccccc|c}
        \toprule
            Method & Param. & sr & gn & ha & pa & or & myv & Avg. \\
            \midrule
                Weight Tuning & 301M & 29.37 &22.14 &31.05 &25.28 &30.17 &28.35 &27.52 \\
                Mask Tuning~\cite{fu2022losses} & 301M & \textbf{25.14} &20.31 &27.62 &22.83 &26.72 &25.99 &24.77 \\
                Adapter Tuning~\cite{le2021lightweight} & 25M & 26.37 &21.16 &28.74 &23.69 &27.52 &27.31 &25.80 \\\cmidrule{1-9}
                Ours & 0.62M & 26.01 &20.72 &28.36 &22.97 &27.04 &26.48 &25.26 \\
                Ours + ft & 301M & 25.23 &\textbf{20.28} &\textbf{27.51} &\textbf{22.75} &\textbf{26.64} &\textbf{25.86} &\textbf{24.71} \\
        \bottomrule
    \end{tabular}
    }
    \vspace{-2em}
    \label{tab:low-resource}
\end{table}

\begin{table}[t]\centering
\vspace{-0.4em}
    \caption{Validating \METHODNS 's multilingual scalability on the multilingual dataset with additional Arabic dialects. }
    \resizebox{\linewidth}{!}{
    \begin{tabular}{c|ccc|ccc}
        \toprule
        Training Data &\multicolumn{3}{c|}{Multilingual w/o ar dialect} &\multicolumn{3}{c}{Multilingual w ar dialect} \\\cmidrule{1-7}
        Split &ar &non-ar &Avg &ar and dialect &non-ar &Avg \\
        \midrule
        Shared Weight Tuning &27.73 &19.93 &20.08 &22.61 &20.21 &20.42 \\
        \cite{fu2022losses} &29.18 &22.77 &22.90 &25.11 &23.02 &23.20 \\
        \cite{pham2022adaptive} &25.24 &17.40 &17.55 &20.14 &17.62 &17.84 \\
        \cite{nguyen2022refining}&23.87 &15.16 &15.33 &18.62 &15.31 &15.60 \\\cmidrule{1-7}
        Ours &23.26 &14.65 &14.81 &18.21 &14.63 &14.94 \\
        \bottomrule
    \end{tabular}
    }
    \vspace{-2em}
    \label{tab:dialect}
\end{table}

\begin{figure*}[!htp]
    \centering
    \vspace{-0.3em}
    \includegraphics[width=\linewidth]{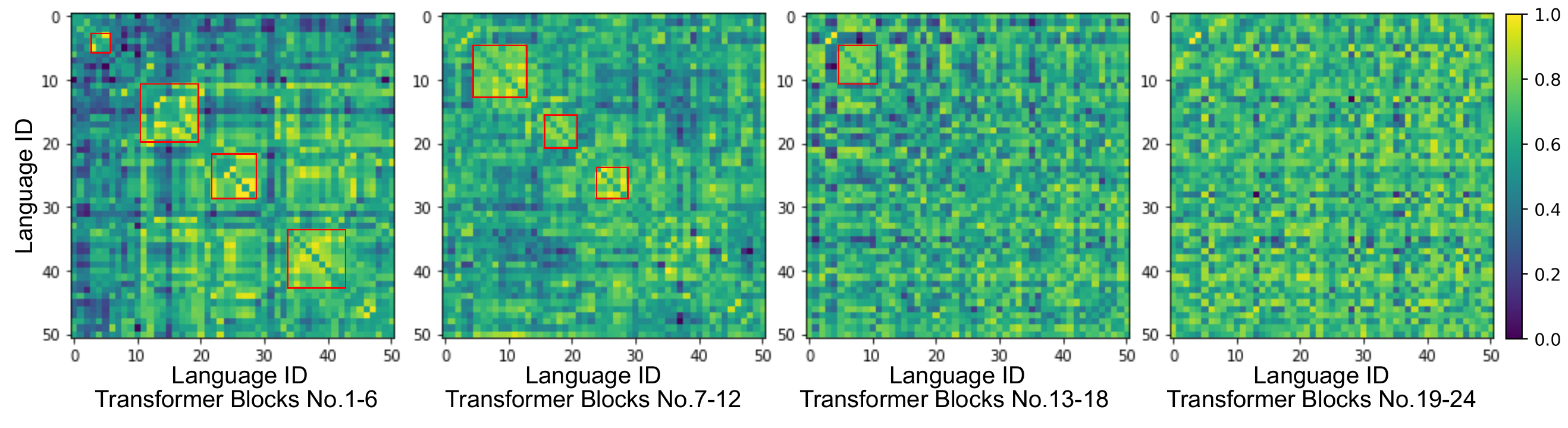}
    \vspace{-2.5em}
    \caption{Visualize the cosine similarity of learned language \MODULE mapping at different transformer blocks. We split the model evenly into four parts, with each part consisting of six transformer blocks. We number these blocks from shallow to deep with 1 to 24. }
    \label{fig:vis_skill}
    \vspace{-1.5em}
\end{figure*}

It is worth noting that when compared with the recently proposed SOTA multilingual ASR system~\cite{nguyen2022refining}, our \METHOD surpasses it in all aspects, including a 0.13 lower CER, 30\% less inference latency, and 16 times less storage overhead. 
We attribute this improvement to the following factors: \uline{(1)} Master-ASR adaptively combines different Specialist Scores for different target languages, while~\cite{nguyen2022refining} uses separate weights for each target language. Such an adaptive sharing approach utilized by Master-ASR could act as a regularizer, encouraging the model to learn features that are more generalizable across languages, resulting in reduced over-fitting on the training data; \uline{(2)} While~\cite{nguyen2022refining} directly updates model weights, Master-ASR incorporates the benefits of mask tuning methods observed in previous works~\cite{fu2022losses}. Specifically, a recent work by~\citet{fu2022losses} has shown that mask tuning can reduce over-fitting, resulting in improved accuracy on ASR tasks. This is achieved by making slight adjustments to the model connections, thereby alleviating the risk of excessively undermining the learned features during pretraining.

\vspace{-0.3em}
Furthermore, we compare \METHOD with its variant \METHODNS -Adapter. As shown in Table~\ref{tab:multilingual}, \METHOD achieves a 3.11 lower CER on average than \METHODNS -Adapter, indicating the superiority of our mask-based design over adapter-based ones.

\begin{table*}[!htp]
    \caption{Validating the trained backbone weight's generalization ability by tuning different backbone weights to different languages.}
    \centering
    \resizebox{\linewidth}{!}{
    \begin{tabular}{cc|cccccc|cccccc}
        \toprule
        \multicolumn{2}{c}{Language} &\multicolumn{6}{c}{Seen languages} &\multicolumn{6}{c}{Unseen low-resource languages} \\
        \midrule
        Method &Backbone &de-1h &zh-1h &es-1h &tt-1h &ru-1h &avg &sr-10m &gn-10m &ha-10m &pa-10m &or-10m &avg \\
        \midrule
        Vanilla Tuning &XLSR &14.45 &21.53 &9.59 &10.11 &14.77 &14.09 &29.37 &22.14 &31.05 &25.28 &30.17 &28.35 \\\cmidrule{1-14}
        \multirow{2}{*}{Mask Tuning} &XLSR &\textbf{14.11} &21.06 &\textbf{9.42} &10.04 &\textbf{14.52} &13.83 &25.14 &20.31 &27.62 &22.83 &\textbf{26.72} &24.52 \\\cmidrule{2-14}
        &Ours &14.12 &\textbf{21.02} &9.45 &\textbf{9.91} &14.58 &\textbf{13.81} &\textbf{25.01} &\textbf{20.17} &\textbf{27.58} &\textbf{22.74} &26.77 &\textbf{24.45} \\
        \bottomrule
    \end{tabular}
    }
    \vspace{-1.7em}
    \label{tab:backbone}
\end{table*}

\textbf{Benchmark on low-resource adaptation.} 
We further evaluate the low-resource tuning performance of \METHOD on six languages, each with only 10 minutes of training data. As shown in Table~\ref{tab:low-resource}, we observe that although \METHOD only adjusts the learned combination of \MODULENS s, it achieves a 0.54$\sim$2.26 lower CER than vanilla weight tuning and adapter tuning~\cite{le2021lightweight}. This indicates that during the training process on our collected multilingual dataset, the learned \MODULENS s in \METHOD extract generalizable features from the training languages, enabling fast adaptation to low-resource languages that have never been seen. 
As such, compared to directly updating the model weights, learning a combination of generalizable features offers strong regularization, avoiding the commonly observed issues of over-fitting and catastrophic forgetting~\cite{winata2020meta,kessler2021continual,hou2021exploiting,cai2014stochastic}. 

\begin{figure}[!tp]
    \centering  
    \includegraphics[width=\linewidth]{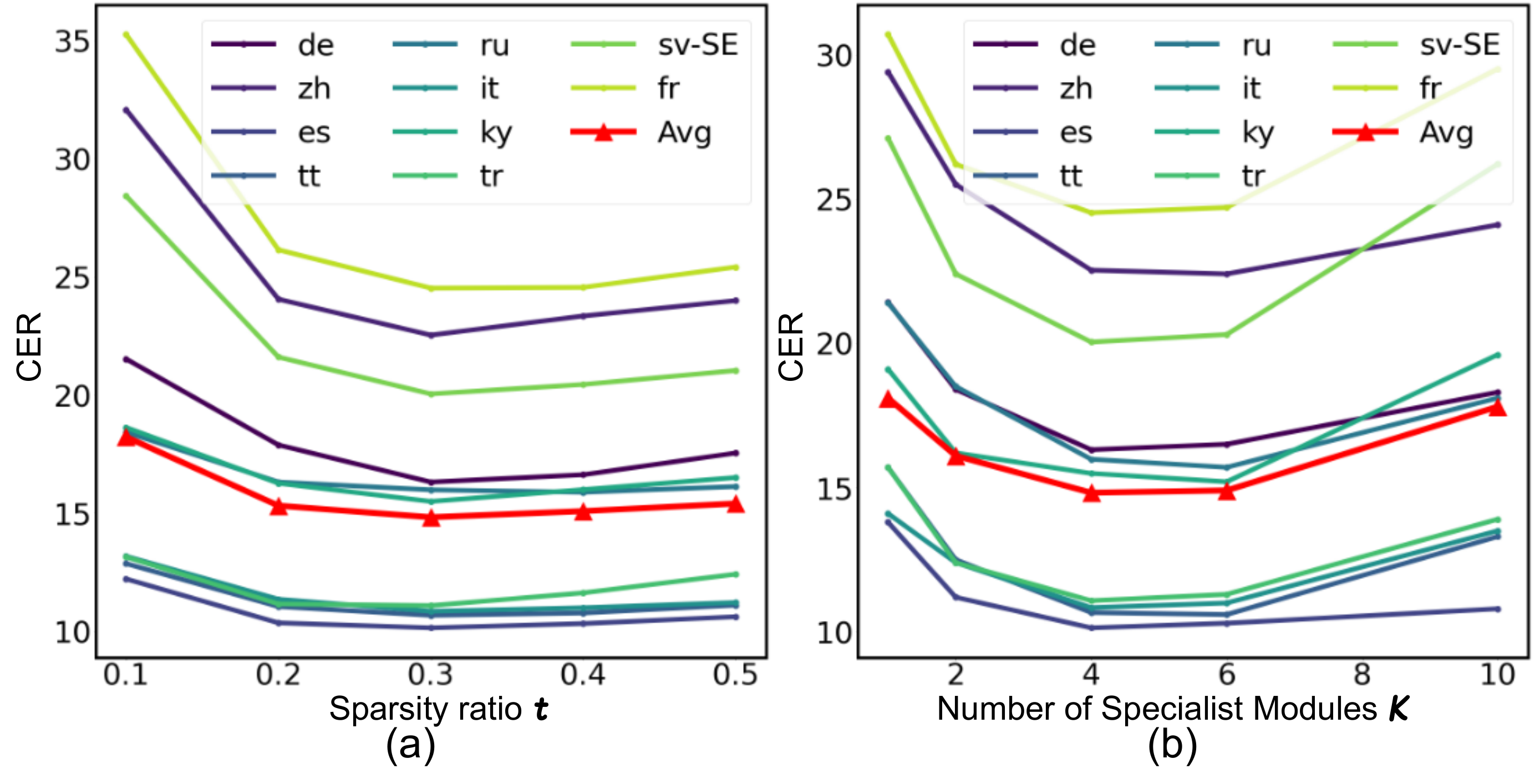}
    \vspace{-2.2em}
    \caption{Validating the achieved CER of \METHOD under different (a) $t$ and (b) $K$ on the multilingual dataset.}
    \label{fig:sparsity}
    \vspace{-2em}
\end{figure}

When compared with the SOTA low-resource tuning method, mask tuning~\cite{fu2022losses}, \METHOD exhibits a slightly worse CER despite reducing the number of trainable parameters by nearly 50 times compared to mask tuning.
We hypothesize that this discrepancy arises because \METHOD encodes generalizable features across multiple languages, leading to decent accuracy in multilingual ASR. However, this generalization capability may compromise the achievable accuracy in a specific language.
In light of this, to further improve \METHODNS 's accuracy on low-resource datasets, we propose to further tune the learned mask for the target low-resource language on top of the tuned \METHODNS. 
We refer to this tuning strategy as \METHODNS +ft. As demonstrated in Table~\ref{tab:low-resource}, \METHODNS +ft attains higher accuracy with an average reduction of 0.06 in CER compared to mask tuning~\cite{fu2022losses}.
This suggests that the learned masks of \METHOD have served as a promising starting point, and incorporating additional mask tuning has the potential to achieve new SOTA accuracy for low-resource adaptation.

\vspace{-0.6em}
\subsection{Case Study: Extending Arabic Dialect}
\vspace{-0.63em}
\textbf{Experiment setup.}
We further add four Arabic dialects~\cite{ali2017speech}, each with 1-hour training data, to our collected multilingual dataset, resulting in a more challenging 56-language dataset. Our aim is to validate whether \METHOD can handle the larger dataset with the more imbalanced data distribution caused by the high similarity between Arabic and its dialects, making a larger portion of data in the training dataset very similar to Arabic. 

\textbf{Observation and analysis.}
As shown in Table~\ref{tab:dialect}, although better accuracy can be achieved on Arabic-related languages after adding Arabic dialects, all methods except \METHOD suffer from a considerable CER increase on non-Arabic languages. We suspect that this is because of the newly added dialects, which are very similar to the standard Arabic we used. The introduction of these dialects results in imbalanced data distribution, making the trained model over-fitted to the Arabic-like features.
Our Master-ASR alleviates this issue by adaptively sharing \MODULENS s, which allocates a moderate number of \MODULENS s for Arabic-related languages and remains others for effectively processing other languages, thus decoupling the learned features for Arabic-related languages and those for other languages.
This further certifies that the proposed \METHOD can handle the more challenging and complex multilingual ASR tasks. 

\vspace{-0.9em}
\subsection{The Impact of Tuning on Generalization Capability}
\vspace{-0.6em}
\citet{fu2022losses} observed that tuning an SSL pretrained model on one specific language will hurt its generalization capability on other languages. To test whether our Master-ASR's training pipeline suffers from a similar issue, we tune the masks on top of the backbone weight learned by \METHOD for different languages and benchmark the achievable CER with those achieved by directly tuning the masks on a pretrained XLSR model. As shown in Table~\ref{tab:backbone}, we observe that \underline{(1)} tuning the masks on \METHODNS 's learned backbone weights consistently leads to lower CER on both seen languages during training and unseen low-resource languages as compared to vanilla weight tuning on a pretrained XLSR model, indicating that Master-ASR's backbone weights can maintain a decent generalization capability; and \underline{(2)} tuning the masks on \METHODNS 's learned backbone weights can even achieve a comparable CER, e.g., -0.06$\sim$ 0.13 lower CER, with mask tuning on the SSL pretrained weight, implying that the training pipeline of \METHOD can preserve or even slightly improve the generalization ability of the backbone weights.

\vspace{-0.9em}
\subsection{Visualizing the Learned Mapping Matrix}
\vspace{-0.6em}
To understand whether \METHOD can leverage the similarity between languages at different levels and adaptively share \MODULENS s across different languages as analyzed in Sec.~\ref{sec:motivate}, we visualize the learned \MODULE combination as an indicator for the captured language similarity. In particular, we number the transformer blocks in the model from 1 to 24, where the larger the number, the deeper the block. We next evenly split the model into four parts with each part consisting of six transformer blocks. For each language $l$, we generate the combination feature $F_l=[T[l,:]]$, $\forall T\in\mathcal{T}_i$, where $\mathcal{T}_i$ is the set of $T$ that belongs to the $i$-th part of the model. We calculate the pairwise cosine similarity between each language pair as shown in Fig.~\ref{fig:vis_skill}, where we reorder the languages based on the language family. The detailed ID and language family for each language can be found in Appendix~\ref{appendix:name}. 

As shown in Fig.~\ref{fig:vis_skill}, we observe that \underline{(1)} \METHOD can successfully learn the similarity among languages. As marked in the red boxes, the  languages belonging to the same language family usually share a higher cosine similarity; and \underline{(2)} different parts of the model behave differently. In particular, a high similarity between language pairs that belong to the same language family is observed in shallow layers; while in deeper layers, languages belonging to the same family show limited similarity since the highly abstract  features learned by the model are less correlated with the human-defined language family.  

\vspace{-0.7em}
\subsection{Ablation Studies}
\vspace{-0.6em}
\textbf{Ablate on sparsity ratio.}
The sparsity ratio $t$ plays a critical role in \METHODNS 's achievable performance. A higher value of $t$ improves inference efficiency but may result in a larger gap in expressive power compared to a dense layer. On the other hand, a smaller value of $t$ incurs higher inference overhead but better preserves the pretrained representation. To this end, we ablate on the impact of $t$ on the achievable CER. As shown in Fig.~\ref{fig:sparsity} (a), we observe that the optimal $t$ is around 0.3, which minimizes the CER across all languages. As compared to the observation in~\cite{fu2022losses}, where the optimal sparsity is around 0.1, we conjecture that the increase in the optimal sparsity is because \METHOD requires more diverse masks to encode generalizable features across different languages, thus necessitating a higher sparsity ratio.

\textbf{Ablate on the number of \MODULENS s in each \LAYERNS.}
The number of \MODULENS s $K$ in each Artisan Layer is a critical parameter for \METHODNS 's achievable CER. Specifically, a larger $K$ may prevent \METHOD from adaptively sharing different \MODULENS s across different languages, while a smaller $K$ can restrict the representation power of \METHODNS , making it unable to learn complex and generalizable features. To this end, we train \METHOD on the multilingual dataset with different values of $K$ and report the results in Fig.~\ref{fig:sparsity} (b). We can observe that \underline{(1)} $K=4$ is the optimal design choice, and \underline{(2)} although excessively large or small $K$ leads to a non-trivial increase in CER, selecting $K$ within a reasonable range (e.g., between 4 to 6) can ensure \METHODNS 's decent accuracy. 

\vspace{-0.2em}
\textbf{Ablate on binary mask generation.}
The way to generate the binary mask from $M_k$ impacts the convergence of our proposed Master-ASR. To validate the effectiveness of our \textit{TopK} selection strategy in binary mask generation, given a Specialist Score $M$, we compare our adopted TopK-based method with two variants including (1) \textit{Thres}, where we generate binary mask $B=\mathbbm{1}_{\sigma(M)<0.5}$ and (2) \textit{Learned}, where we introduce a learnable threshold value $\theta$ and generate the binary mask $B=\mathbbm{1}_{B=M<\theta}$. As shown in Table~\ref{tab:mask_gen}, our adopted TopK-based mask generation shows the best recognition accuracy, e.g., a 4.12 and 10.97 lower CER over \textit{Thres} and \textit{Learned}, respectively. This indicates that properly selecting the binary mask generation method is crucial to the achievable accuracy and our TopK-based method is a decent design choice for our Master-ASR framework.

\begin{table}[]
    \centering
    \caption{Ablate on the binary mask generation scheme.}
        \vspace{0.5em}
    \resizebox{0.6\linewidth}{!}{
    \begin{tabular}{c|ccc}
    \toprule
       Method & \textit{Thres} & \textit{Learned} & \textit{\textbf{TopK}} \\
       \midrule
       CER & 18.36 & 25.21 & \textbf{14.24} \\
       \bottomrule
    \end{tabular}
    }
    \vspace{-2em}
    \label{tab:mask_gen}
\end{table}

\vspace{-0.2em}
\textbf{Ablate on the weight update scheme.}
One of the key differences between Master-ASR and existing mask-based ASR tuning frameworks~\cite{fu2022losses,lai2021parp} is the adopted weight update scheme. To understand to what extent different weight update methods can impact the achievable ASR performance, we conduct an ablation study to compare our adopted iterative update (i.e., \textit{Iter}) with two variants (1) \textit{Freeze}, where the model weight is frozen throughout the whole tuning process, (2) \textit{Random}, where we randomly select $k$ elements to update (e.g., $k$ as defined in Sec.~\ref{sec:model}). As shown in Table~\ref{tab:weight_update}, we observe that \underline{(1)} as long as the model weight can be updated during tuning to accommodate the learned binary mask, the multilingual ASR system can achieve promising performance, e.g., \textit{Random} and \textit{Iter} achieve 4.06 and 4.23 lower CER than \textit{Freeze}, respectively, and \underline{(2)} our dedicated weight update policy, i.e., \textit{Iter}, can further reduce the achievable CER, e.g., a 0.17 reduction as compared to \textit{Random}.

\vspace{-1.2em}
\section{Limitations and Future Directions}
\vspace{-0.6em}
Despite the promising performance achieved by our proposed Master-ASR, there are still several directions that can further improve Master-ASR's multilingual and low-resource performance that are worth further exploration. Here, we list a few of them:
\vspace{-1.2em}
\begin{itemize}
    \item \textbf{Learning a set of more representative and generalizable Specialist Scores.} While we observe promising performance on multilingual and low-resource ASR by simply learning a set of Specialist Scores in Master-ASR, these modules may not provide sufficiently generalizable features for better tuning low-resource languages. One potential solution is integrating techniques like contrastive learning to help Master-ASR learn a more generalizable set of Specialist Scores.
    \vspace{-0.9em}
    \item \textbf{Adaptively introducing new Specialist Scores during tuning.} When the targeting low-resource language has a significant distribution shift with the training languages, the commonly used generalizable Specialist Scores may not sufficiently fit the new distribution. Thus, adaptively introducing a new Specialist Score in the model to better accommodate the significant distribution shift, in this case, may further improve the performance.
    \vspace{-0.9em}
    \item \textbf{Guide tuning with prior knowledge.} Some existing works~\cite{zhao2020low,li2020towards} show that using prior knowledge about languages can help the model to make better decisions. Master-ASR may also benefit from incorporating prior knowledge to guide the tuning process, especially under the low-resource scenario. Specifically, it is worth exploring whether human-defined language families can help to generate a combination strategy for Specialist Scores, even without the need to further tune the model. 
    \vspace{-0.9em}
\end{itemize}



\begin{table}[t]
    \centering
    \caption{Ablate on the weight update strategy. }
    \vspace{0.5em}
    \resizebox{0.6\linewidth}{!}{
    \begin{tabular}{c|ccc}
    \toprule
        Method & \textit{Freeze} & \textit{Random} & \textit{\textbf{Iter}} \\
        \midrule
        CER & 18.47 & 14.41 & 14.24 \\
        \bottomrule
    \end{tabular}
    }
    \vspace{-2em}
    \label{tab:weight_update}
\end{table}
\vspace{-1.2em}
\section{Conclusion}
\vspace{-0.5em}
This work presents an ASR framework, dubbed \METHODNS. To the best of our knowledge, \METHOD is the first that can  simultaneously achieve strong multilingual scalability and low-resource adaptation ability in ASR thanks to its modularize-then-assemble strategy. Specifically, \METHOD learns a set of generalizable \MODULENS   
 s and adaptively assembles them for different languages to reduce the multilingual overhead and enable effective knowledge transfer for low-resource adaptation. 
 Extensive experiments consistently validate the effectiveness of \METHOD in boosting the scalability and low-resource adaptation capability of ASR models. For example, (1) in multilingual ASR, \METHOD achieves a 0.13$\sim$2.41 lower CER with 30\% smaller inference overhead over SOTA ASR methods; (2) in low-resource tuning, \METHOD achieves a comparable CER with nearly 50 times fewer trainable parameters over SOTA ASR methods.

\vspace{-1.3em}
\section*{Acknowledgements}
\vspace{-0.6em}
This work was supported in part by CoCoSys, one of the seven centers in JUMP 2.0, a Semiconductor Research Corporation (SRC) program sponsored by DARPA and an IBM faculty award received by Dr. Yingyan (Celine) Lin.

\bibliography{icml2023}

\begin{thebibliography}{47}
\providecommand{\natexlab}[1]{#1}
\providecommand{\url}[1]{\texttt{#1}}
\expandafter\ifx\csname urlstyle\endcsname\relax
  \providecommand{\doi}[1]{doi: #1}\else
  \providecommand{\doi}{doi: \begingroup \urlstyle{rm}\Url}\fi

\bibitem[Ali et~al.(2017)Ali, Vogel, and Renals]{ali2017speech}
Ali, A., Vogel, S., and Renals, S.
\newblock Speech recognition challenge in the wild: Arabic mgb-3.
\newblock In \emph{2017 IEEE Automatic Speech Recognition and Understanding
  Workshop (ASRU)}, pp.\  316--322. IEEE, 2017.

\bibitem[Ali et~al.(1999)Ali, Van~der Spiegel, Mueller, Haentjens, and
  Berman]{ali1999acoustic}
Ali, A.~A., Van~der Spiegel, J., Mueller, P., Haentjens, G., and Berman, J.
\newblock An acoustic-phonetic feature-based system for automatic phoneme
  recognition in continuous speech.
\newblock In \emph{1999 IEEE International Symposium on Circuits and Systems
  (ISCAS)}, volume~3, pp.\  118--121. IEEE, 1999.

\bibitem[Ao et~al.(2021)Ao, Wang, Zhou, Liu, Ren, Wu, Ko, Li, Zhang, Wei,
  et~al.]{ao2021speecht5}
Ao, J., Wang, R., Zhou, L., Liu, S., Ren, S., Wu, Y., Ko, T., Li, Q., Zhang,
  Y., Wei, Z., et~al.
\newblock Speecht5: Unified-modal encoder-decoder pre-training for spoken
  language processing.
\newblock \emph{arXiv preprint arXiv:2110.07205}, 2021.

\bibitem[Ardila et~al.(2019)Ardila, Branson, Davis, Henretty, Kohler, Meyer,
  Morais, Saunders, Tyers, and Weber]{ardila2019common}
Ardila, R., Branson, M., Davis, K., Henretty, M., Kohler, M., Meyer, J.,
  Morais, R., Saunders, L., Tyers, F.~M., and Weber, G.
\newblock Common voice: A massively-multilingual speech corpus.
\newblock \emph{arXiv preprint arXiv:1912.06670}, 2019.

\bibitem[Babu et~al.(2021)Babu, Wang, Tjandra, Lakhotia, Xu, Goyal, Singh, von
  Platen, Saraf, Pino, et~al.]{babu2021xls}
Babu, A., Wang, C., Tjandra, A., Lakhotia, K., Xu, Q., Goyal, N., Singh, K.,
  von Platen, P., Saraf, Y., Pino, J., et~al.
\newblock Xls-r: Self-supervised cross-lingual speech representation learning
  at scale.
\newblock \emph{arXiv preprint arXiv:2111.09296}, 2021.

\bibitem[Baevski et~al.(2020)Baevski, Zhou, Mohamed, and
  Auli]{baevski2020wav2vec}
Baevski, A., Zhou, Y., Mohamed, A., and Auli, M.
\newblock wav2vec 2.0: A framework for self-supervised learning of speech
  representations.
\newblock \emph{Advances in Neural Information Processing Systems},
  33:\penalty0 12449--12460, 2020.

\bibitem[Cai et~al.(2014)Cai, Shi, and Liu]{cai2014stochastic}
Cai, M., Shi, Y., and Liu, J.
\newblock Stochastic pooling maxout networks for low-resource speech
  recognition.
\newblock In \emph{2014 IEEE International Conference on Acoustics, Speech and
  Signal Processing (ICASSP)}, pp.\  3266--3270. IEEE, 2014.

\bibitem[Campbell(2008)]{campbell2008ethnologue}
Campbell, L.
\newblock Ethnologue: Languages of the world, 2008.

\bibitem[Cao et~al.(2022)Cao, Prakash, and Hamza]{Cao2022AttentionFA}
Cao, J., Prakash, C., and Hamza, W.
\newblock Attention fusion: a light yet efficient late fusion mechanism for
  task adaptation in nlu.
\newblock In \emph{NAACL-HLT}, 2022.

\bibitem[Chen \& Mak(2015)Chen and Mak]{Chen2015MultitaskLO}
Chen, D. and Mak, B. K.-W.
\newblock Multitask learning of deep neural networks for low-resource speech
  recognition.
\newblock \emph{IEEE/ACM Transactions on Audio, Speech, and Language
  Processing}, 23:\penalty0 1172--1183, 2015.

\bibitem[Conneau et~al.(2020)Conneau, Baevski, Collobert, Mohamed, and
  Auli]{conneau2020unsupervised}
Conneau, A., Baevski, A., Collobert, R., Mohamed, A., and Auli, M.
\newblock Unsupervised cross-lingual representation learning for speech
  recognition.
\newblock \emph{arXiv preprint arXiv:2006.13979}, 2020.

\bibitem[Crawshaw(2020)]{Crawshaw2020MultiTaskLW}
Crawshaw, M.
\newblock Multi-task learning with deep neural networks: A survey.
\newblock \emph{ArXiv}, abs/2009.09796, 2020.

\bibitem[Deligne et~al.(2001)Deligne, Eide, and
  Gopinath]{Deligne2001LowResourceSR}
Deligne, S., Eide, E., and Gopinath, R.~A.
\newblock Low-resource speech recognition of 500-word vocabularies.
\newblock 2001.

\bibitem[Du et~al.(2022)Du, Zhang, shi Zhu, Dai, Wu, Fang, and Yang]{Du2022ACJ}
Du, Y., Zhang, J., shi Zhu, Q., Dai, L., Wu, M., Fang, X., and Yang, Z.-W.
\newblock A complementary joint training approach using unpaired speech and
  text for low-resource automatic speech recognition.
\newblock \emph{ArXiv}, abs/2204.02023, 2022.

\bibitem[Finn et~al.(2017)Finn, Abbeel, and Levine]{Finn2017ModelAgnosticMF}
Finn, C., Abbeel, P., and Levine, S.
\newblock Model-agnostic meta-learning for fast adaptation of deep networks.
\newblock In \emph{International Conference on Machine Learning}, 2017.

\bibitem[Fu et~al.(2022)Fu, Zhang, Qian, Ye, Yu, Lai, and Lin]{fu2022losses}
Fu, Y., Zhang, Y., Qian, K., Ye, Z., Yu, Z., Lai, C.-I., and Lin, Y.
\newblock Losses can be blessings: Routing self-supervised speech
  representations towards efficient multilingual and multitask speech
  processing.
\newblock \emph{arXiv preprint arXiv:2211.01522}, 2022.

\bibitem[Graves et~al.(2013)Graves, Jaitly, and Mohamed]{graves2013hybrid}
Graves, A., Jaitly, N., and Mohamed, A.-r.
\newblock Hybrid speech recognition with deep bidirectional lstm.
\newblock In \emph{2013 IEEE workshop on automatic speech recognition and
  understanding}, pp.\  273--278. IEEE, 2013.

\bibitem[Heafield et~al.(2013)Heafield, Pouzyrevsky, Clark, and
  Koehn]{heafield2013scalable}
Heafield, K., Pouzyrevsky, I., Clark, J.~H., and Koehn, P.
\newblock Scalable modified kneser-ney language model estimation.
\newblock In \emph{Proceedings of the 51st Annual Meeting of the Association
  for Computational Linguistics (Volume 2: Short Papers)}, pp.\  690--696,
  2013.

\bibitem[Hou et~al.(2021)Hou, Zhu, Wang, Wang, Qin, Xu, and
  Shinozaki]{hou2021exploiting}
Hou, W., Zhu, H., Wang, Y., Wang, J., Qin, T., Xu, R., and Shinozaki, T.
\newblock Exploiting adapters for cross-lingual low-resource speech
  recognition.
\newblock \emph{IEEE/ACM Transactions on Audio, Speech, and Language
  Processing}, 30:\penalty0 317--329, 2021.

\bibitem[Hsu et~al.(2020)Hsu, Chen, and Lee]{hsu2020meta}
Hsu, J.-Y., Chen, Y.-J., and Lee, H.-y.
\newblock Meta learning for end-to-end low-resource speech recognition.
\newblock In \emph{ICASSP 2020-2020 IEEE International Conference on Acoustics,
  Speech and Signal Processing (ICASSP)}, pp.\  7844--7848. IEEE, 2020.

\bibitem[Hsu et~al.(2021)Hsu, Bolte, Tsai, Lakhotia, Salakhutdinov, and
  Mohamed]{hsu2021hubert}
Hsu, W.-N., Bolte, B., Tsai, Y.-H.~H., Lakhotia, K., Salakhutdinov, R., and
  Mohamed, A.
\newblock Hubert: Self-supervised speech representation learning by masked
  prediction of hidden units.
\newblock \emph{IEEE/ACM Transactions on Audio, Speech, and Language
  Processing}, 29:\penalty0 3451--3460, 2021.

\bibitem[Hu et~al.(2021)Hu, Shen, Wallis, Allen-Zhu, Li, Wang, Wang, and
  Chen]{hu2021lora}
Hu, E.~J., Shen, Y., Wallis, P., Allen-Zhu, Z., Li, Y., Wang, S., Wang, L., and
  Chen, W.
\newblock Lora: Low-rank adaptation of large language models.
\newblock \emph{arXiv preprint arXiv:2106.09685}, 2021.

\bibitem[Jaiswal et~al.(2022)Jaiswal, Ma, Chen, Ding, and
  Wang]{jaiswal2022training}
Jaiswal, A.~K., Ma, H., Chen, T., Ding, Y., and Wang, Z.
\newblock Training your sparse neural network better with any mask.
\newblock In \emph{International Conference on Machine Learning}, pp.\
  9833--9844. PMLR, 2022.

\bibitem[Kahn et~al.(2019)Kahn, Lee, and Hannun]{Kahn2019SelfTrainingFE}
Kahn, J., Lee, A., and Hannun, A.~Y.
\newblock Self-training for end-to-end speech recognition.
\newblock \emph{ICASSP 2020 - 2020 IEEE International Conference on Acoustics,
  Speech and Signal Processing (ICASSP)}, pp.\  7084--7088, 2019.

\bibitem[Kessler et~al.(2021)Kessler, Thomas, and Karout]{kessler2021continual}
Kessler, S., Thomas, B., and Karout, S.
\newblock Continual-wav2vec2: an application of continual learning for
  self-supervised automatic speech recognition.
\newblock \emph{arXiv preprint arXiv:2107.13530}, 2021.

\bibitem[Kirsch et~al.(2018)Kirsch, Kunze, and Barber]{Kirsch2018ModularNL}
Kirsch, L., Kunze, J., and Barber, D.
\newblock Modular networks: Learning to decompose neural computation.
\newblock In \emph{Neural Information Processing Systems}, 2018.

\bibitem[Lai et~al.(2021)Lai, Zhang, Liu, Chang, Liao, Chuang, Qian, Khurana,
  Cox, and Glass]{lai2021parp}
Lai, C.-I.~J., Zhang, Y., Liu, A.~H., Chang, S., Liao, Y.-L., Chuang, Y.-S.,
  Qian, K., Khurana, S., Cox, D., and Glass, J.
\newblock Parp: Prune, adjust and re-prune for self-supervised speech
  recognition.
\newblock \emph{Advances in Neural Information Processing Systems},
  34:\penalty0 21256--21272, 2021.

\bibitem[Le et~al.(2021)Le, Pino, Wang, Gu, Schwab, and
  Besacier]{le2021lightweight}
Le, H., Pino, J., Wang, C., Gu, J., Schwab, D., and Besacier, L.
\newblock Lightweight adapter tuning for multilingual speech translation.
\newblock \emph{arXiv preprint arXiv:2106.01463}, 2021.

\bibitem[Li et~al.(2019)Li, Sainath, Pang, and Wu]{Li2019SemisupervisedTF}
Li, B., Sainath, T.~N., Pang, R., and Wu, Z.
\newblock Semi-supervised training for end-to-end models via weak distillation.
\newblock \emph{ICASSP 2019 - 2019 IEEE International Conference on Acoustics,
  Speech and Signal Processing (ICASSP)}, pp.\  2837--2841, 2019.

\bibitem[Li et~al.(2021)Li, Pang, Sainath, Gulati, Zhang, Qin, Haghani, Huang,
  Ma, and Bai]{li2021scaling}
Li, B., Pang, R., Sainath, T.~N., Gulati, A., Zhang, Y., Qin, J., Haghani, P.,
  Huang, W.~R., Ma, M., and Bai, J.
\newblock Scaling end-to-end models for large-scale multilingual asr.
\newblock In \emph{2021 IEEE Automatic Speech Recognition and Understanding
  Workshop (ASRU)}, pp.\  1011--1018. IEEE, 2021.

\bibitem[Li et~al.(2022)Li, Pang, Zhang, Sainath, Strohman, Haghani, Zhu,
  Farris, Gaur, and Prasad]{li2022massively}
Li, B., Pang, R., Zhang, Y., Sainath, T.~N., Strohman, T., Haghani, P., Zhu,
  Y., Farris, B., Gaur, N., and Prasad, M.
\newblock Massively multilingual asr: A lifelong learning solution.
\newblock In \emph{ICASSP 2022-2022 IEEE International Conference on Acoustics,
  Speech and Signal Processing (ICASSP)}, pp.\  6397--6401. IEEE, 2022.

\bibitem[Li et~al.(2020)Li, Dalmia, Mortensen, Li, Black, and
  Metze]{li2020towards}
Li, X., Dalmia, S., Mortensen, D., Li, J., Black, A., and Metze, F.
\newblock Towards zero-shot learning for automatic phonemic transcription.
\newblock In \emph{Proceedings of the AAAI Conference on Artificial
  Intelligence}, volume~34, pp.\  8261--8268, 2020.

\bibitem[Liang et~al.(2020)Liang, Wu, Ziyin, Morency, and
  Salakhutdinov]{liang2020learning}
Liang, P.~P., Wu, P., Ziyin, L., Morency, L.-P., and Salakhutdinov, R.
\newblock Learning in low-resource modalities via cross-modal generalization.
\newblock 2020.

\bibitem[Liu et~al.(2022)Liu, Tian, Chen, and Shen]{liu2022don}
Liu, S., Tian, Y., Chen, T., and Shen, L.
\newblock Don't be so dense: Sparse-to-sparse gan training without sacrificing
  performance.
\newblock \emph{arXiv preprint arXiv:2203.02770}, 2022.

\bibitem[Miao et~al.(2013)Miao, Metze, and Rawat]{Miao2013DeepMN}
Miao, Y., Metze, F., and Rawat, S.
\newblock Deep maxout networks for low-resource speech recognition.
\newblock \emph{2013 IEEE Workshop on Automatic Speech Recognition and
  Understanding}, pp.\  398--403, 2013.

\bibitem[Nguyen et~al.(2022)Nguyen, Joty, Kui, and Aw]{nguyen2022refining}
Nguyen, X.-P., Joty, S., Kui, W., and Aw, A.~T.
\newblock Refining low-resource unsupervised translation by language
  disentanglement of multilingual model.
\newblock \emph{arXiv preprint arXiv:2205.15544}, 2022.

\bibitem[Nichol et~al.(2018)Nichol, Achiam, and Schulman]{Nichol2018OnFM}
Nichol, A., Achiam, J., and Schulman, J.
\newblock On first-order meta-learning algorithms.
\newblock \emph{ArXiv}, abs/1803.02999, 2018.

\bibitem[Pan \& Rajan(2020)Pan and Rajan]{Pan2020OnDA}
Pan, R. and Rajan, H.
\newblock On decomposing a deep neural network into modules.
\newblock \emph{Proceedings of the 28th ACM Joint Meeting on European Software
  Engineering Conference and Symposium on the Foundations of Software
  Engineering}, 2020.

\bibitem[Pham et~al.(2022)Pham, Waibel, and Niehues]{pham2022adaptive}
Pham, N.-Q., Waibel, A., and Niehues, J.
\newblock Adaptive multilingual speech recognition with pretrained models.
\newblock \emph{arXiv preprint arXiv:2205.12304}, 2022.

\bibitem[Ponti et~al.(2022)Ponti, Sordoni, and Reddy]{Ponti2022CombiningMS}
Ponti, E., Sordoni, A., and Reddy, S.
\newblock Combining modular skills in multitask learning.
\newblock \emph{ArXiv}, abs/2202.13914, 2022.

\bibitem[Pratap et~al.(2020)Pratap, Sriram, Tomasello, Hannun, Liptchinsky,
  Synnaeve, and Collobert]{pratap2020massively}
Pratap, V., Sriram, A., Tomasello, P., Hannun, A., Liptchinsky, V., Synnaeve,
  G., and Collobert, R.
\newblock Massively multilingual asr: 50 languages, 1 model, 1 billion
  parameters.
\newblock \emph{arXiv preprint arXiv:2007.03001}, 2020.

\bibitem[Toshniwal et~al.(2018)Toshniwal, Sainath, Weiss, Li, Moreno,
  Weinstein, and Rao]{toshniwal2018multilingual}
Toshniwal, S., Sainath, T.~N., Weiss, R.~J., Li, B., Moreno, P., Weinstein, E.,
  and Rao, K.
\newblock Multilingual speech recognition with a single end-to-end model.
\newblock In \emph{2018 IEEE international conference on acoustics, speech and
  signal processing (ICASSP)}, pp.\  4904--4908. IEEE, 2018.

\bibitem[Winata et~al.(2020)Winata, Cahyawijaya, Lin, Liu, Xu, and
  Fung]{winata2020meta}
Winata, G.~I., Cahyawijaya, S., Lin, Z., Liu, Z., Xu, P., and Fung, P.
\newblock Meta-transfer learning for code-switched speech recognition.
\newblock \emph{arXiv preprint arXiv:2004.14228}, 2020.

\bibitem[Yadav \& Sitaram(2022)Yadav and Sitaram]{yadav2022survey}
Yadav, H. and Sitaram, S.
\newblock A survey of multilingual models for automatic speech recognition.
\newblock \emph{arXiv preprint arXiv:2202.12576}, 2022.

\bibitem[Zhao et~al.(2020)Zhao, Wu, Tao, Xu, Zhao, and Yan]{zhao2020low}
Zhao, X., Wu, W., Tao, C., Xu, C., Zhao, D., and Yan, R.
\newblock Low-resource knowledge-grounded dialogue generation.
\newblock \emph{arXiv preprint arXiv:2002.10348}, 2020.

\bibitem[Zheng et~al.(2021)Zheng, Xiao, Gong, Zhou, Liang, and
  Lin]{zheng2021wav}
Zheng, G., Xiao, Y., Gong, K., Zhou, P., Liang, X., and Lin, L.
\newblock Wav-bert: Cooperative acoustic and linguistic representation learning
  for low-resource speech recognition.
\newblock \emph{arXiv preprint arXiv:2109.09161}, 2021.

\bibitem[Zhu et~al.(2021)Zhu, An, Zheng, and Ou]{zhu2021multilingual}
Zhu, C., An, K., Zheng, H., and Ou, Z.
\newblock Multilingual and crosslingual speech recognition using
  phonological-vector based phone embeddings.
\newblock In \emph{2021 IEEE Automatic Speech Recognition and Understanding
  Workshop (ASRU)}, pp.\  1034--1041. IEEE, 2021.

\end{thebibliography}
\bibliographystyle{icml2023}

\newpage
\appendix

\section{\METHODNS 's Performance on All Languages}
\label{app:all-lang}
To provide a thorough overview on the performance of our proposed \METHOD with vanilla weight tuning on each langauge separately and two SOTA multilingual ASR solutions~\cite{pham2022adaptive,nguyen2022refining}, we report the achieved CER on each language in Table~\ref{tab:all-lang}. We observe that \METHODNS 's improvement over the SOTA solutions is consistent on most of the languages. 
\begin{table}[!htp]\centering
\caption{The achieved CER on each language in the multilingual ASR dataset from proposed \METHOD and baseline methods.}\label{tab:all-lang}
\resizebox{\linewidth}{!}{
\begin{tabular}{c|ccc|c}\toprule
Language &Separate Weight Tuning & \cite{pham2022adaptive} & \cite{nguyen2022refining}& Ours \\
\midrule
ab &14.25 &19.01 &17.04 &16.81 \\
ar &21.96 &27.14 &24.72 &24.34 \\
ba &11.92 &15.38 &13.50 &13.42 \\
be &10.72 &14.57 &12.88 &12.44 \\
bn &17.94 &23.19 &20.91 &20.64 \\
ca &11.30 &15.04 &13.35 &12.95 \\
ckb &11.83 &15.29 &13.55 &13.58 \\
cs &13.80 &19.51 &16.16 &16.21 \\
cy &10.76 &14.13 &12.20 &11.9 \\
de &14.45 &19.34 &16.77 &16.31 \\
dv &13.25 &17.08 &15.44 &15.35 \\
el &11.46 &14.76 &13.18 &13.01 \\
en &21.20 &27.52 &24.19 &24.06 \\
eo &5.77 &7.82 &6.21 &6.14 \\
es &9.59 &12.21 &10.71 &10.67 \\
et &11.69 &15.27 &13.00 &12.85 \\
eu &6.43 &8.61 &7.43 &7.4 \\
fa &18.87 &24.70 &21.37 &21.21 \\
fr &22.63 &29.37 &25.81 &25.52 \\
fy-NL &8.32 &10.88 &9.41 &9.42 \\
gl &5.47 &6.93 &6.19 &6.17 \\
hi &16.58 &21.57 &18.07 &18.04 \\
hu &12.74 &15.89 &14.05 &13.94 \\
ia &3.69 &5.34 &4.04 &3.99 \\
id &8.59 &11.21 &9.53 &9.42 \\
it &9.99 &12.64 &11.31 &10.84 \\
kab &20.10 &26.46 &22.42 &22.53 \\
kmr &9.61 &12.16 &10.24 &10.31 \\
ky &14.62 &19.04 &15.54 &15.49 \\
lg &10.37 &14.82 &12.14 &12.1 \\
lt &10.10 &14.03 &11.85 &11.78 \\
mhr &8.90 &11.72 &9.61 &9.56 \\
mn &16.55 &22.09 &18.38 &18.34 \\
mr &12.49 &16.33 &13.92 &13.88 \\
nl &7.72 &9.94 &8.31 &8.25 \\
pl &8.39 &12.14 &10.06 &9.88 \\
pt &12.76 &16.68 &14.41 &14.36 \\
ro &7.74 &10.03 &8.44 &8.45 \\
ru &14.77 &19.84 &16.59 &15.98 \\
rw &22.63 &31.29 &25.88 &26.01 \\
sk &18.16 &23.10 &20.25 &20.31 \\
sv-SE &18.21 &24.56 &20.86 &20.54 \\
sw &8.12 &10.59 &8.98 &8.84 \\
ta &12.59 &17.40 &14.59 &14.46 \\
tr &10.33 &13.23 &11.48 &11.08 \\
tt &10.11 &12.21 &10.71 &10.67 \\
ug &9.39 &12.02 &10.60 &10.52 \\
uk &12.36 &15.73 &13.85 &13.78 \\
ur &15.38 &19.40 &17.74 &17.61 \\
uz &9.98 &13.67 &11.23 &11.29 \\
zh &21.53 &26.39 &23.94 &23.53 \\
\bottomrule
\end{tabular}
}
\end{table}

\section{Ablate on Updating $\mathcal{T}$ Effectively}
To validate the effectiveness of our proposed method in Sec.~\ref{sec:training} to ease the training of $\mathcal{T}$, we ablate the impact of $\alpha$ and $\beta$ selections to the trained performance. As shown in Table.~\ref{tab:alpha}, we observe that a larger $\alpha$ significantly helps with training by as high as 4.15 CER reduction. For $\beta$, as long as a larger value (e.g., $\geq$ 5) is selected, \METHOD can consistently achieve better CER. 

\begin{table}[!htp]\centering
\caption{Ablate on $\alpha$ and $\beta$'s impact on the trained multilingual ASR's average CER across all the supported languages. }\label{tab:alpha}
\resizebox{\linewidth}{!}{
\begin{tabular}{c|ccc|ccc}
\toprule
&\multicolumn{3}{c}{a=10} &\multicolumn{3}{c}{b=5} \\\cmidrule{2-7}
&b=1 &b=5 &b=20 &a=0.1 &a=1 &a=10 \\\cmidrule{2-7}
CER &16.21 &14.24 &14.33 &18.39 &17.35 &14.24 \\
\bottomrule
\end{tabular}
}
\end{table}

\section{Ablate on $\mathcal{W}$ and $\mathcal{M}$ Training}
To validate if our proposed pipeline on iterative update $\mathcal{W}$ and $\mathcal{M}$ in Sec.~\ref{sec:training}, we study the different update policies. Specifically, we consider two update scenarios: (1) only update $\mathcal{M}$ without touching $\mathcal{W}$, dubbed $\mathcal{M}$ only, and (2) update $\mathcal{W}$ and $\mathcal{M}$ with different interval $\gamma$. As shown in Table~\ref{tab:mask-convergence}, we observe that unlike the observation in~\citet{fu2022losses}, only tuning $\mathcal{M}$ (i.e., $\mathcal{M}$ only) leads to failure in training with much higher CER than an iterative update. On the other hand, iteratively updating with $\gamma$ ranging from 1000 to 5000 all lead to decent CER, while updating excessively frequently makes the $\mathcal{M}$ and $\mathcal{W}$ cannot be updated enough iterations to fit each other and too infrequent consumes the training iterations optimizing on an already saturated objective, both lead to higher CER. 

\begin{table}[!htp]\centering
\caption{Ablate on different strategies to train $\mathcal{W}$ and $\mathcal{M}$. }\label{tab: }
\resizebox{\linewidth}{!}{
\begin{tabular}{c|c|cccc}\toprule
&$\mathcal{M}$ only & $\gamma = 100$ & $\gamma = 1000$ & $\gamma = 5000$ & $\gamma = 50000$ \\\cmidrule{2-6}
CER &27.52 &18.95 &14.47 &14.24 &22.06 \\
\bottomrule
\end{tabular}
}
    \label{tab:mask-convergence}
\end{table}

\section{Details on Language ID and Family}
\label{appendix:name}
Please refer to next page. 
\begin{table*}[!htp]
\caption{The detailed information on the languages in our multilingual dataset. }\label{tab:multi-name}
\centering
\resizebox{\linewidth}{!}{
\begin{tabular}{c|cccccccccc}\toprule
ID &1 &2 &3 &4 &5 &6 &7 &8 &9 &10 \\
\midrule
Language &Welsh &English &German &Frisian &Dutch &Swedish &Catalan &French &Spanish &Italian \\
Abbreviation &cy &en &de &fy-NL &nl &sv-SE &ca &fr &es &it \\
Language Family &Indo-European &Indo-European &Indo-European &Indo-European &Indo-European &Indo-European &Indo-European &Indo-European &Indo-European &Indo-European \\
\midrule
ID &11 &12 &13 &14 &15 &16 &17 &18 &19 &20 \\
\midrule
Language &Portuguese &Romanian &Galician &Belarusian &Russian &Polish &Czech &Ukrainian &Slovak &Lithuanian \\
Abbreviation &pt &ro &gl &be &ru &pl &cs &uk &sk \\
Language Family &Indo-European &Indo-European &Indo-European &Indo-European &Indo-European &Indo-European &Indo-European &Indo-European &Indo-European &Indo-European \\
\midrule
ID &21 &22 &23 &24 &25 &26 &27 &28 &29 &30 \\
\midrule
Language &Greek &Persian &Bengali &Dhivehi &Central Kurdish &Kurmanji Kurdish &Urdu &Marathi &Hindi &Mandrian \\
Abbreviation &el &fa &bn &dv &ckb &kmr &ur &mr &hi &zh \\
Language Family &Indo-European &Indo-European &Indo-European &Indo-European &Indo-European &Indo-European &Indo-European &Indo-European &Indo-European &Sino-Tibetan \\
\midrule
ID &31 &32 &33 &34 &35 &36 &37 &38 &39 &40 \\
\midrule
Language &Arabic &Kabyle &Meadow Mari &Estonian &Hungarian &Bashkir &Uzbek &Turkish &Uyghur &Kyrgyz \\
Abbreviation &ar &kab &mhr &et &hu &ba &uz &tr &ug &ky \\
Language Family &Semitic-Hamitic &Semitic-Hamitic &Uralic &Uralic &Uralic &Altaic &Altaic &Altaic &Altaic &Altaic \\
\midrule
ID &41 &42 &43 &44 &45 &46 &47 &48 &49 &50 \\
\midrule
Language &Tatar &Mongolian &Abkhaz &Indonesian &Tamil &Kinyarwanda &Luganda &Swahili &Esperanto &Basque \\
Abbreviation &tt &mn &ab &id &ta &rw &lg &sw &eo &eu \\
Language Family &Altaic &Altaic &Caucasian &Austronesian &Dravidian &Niger-Congo &Niger-Congo &Niger-Congo &Other &Other \\
\midrule
ID &51 & & & & & & & & & \\
\midrule
Language &Interlingua & & & & & & & & & \\
Abbreviation &ia & & & & & & & & & \\
Language Family &Other & & & & & & & & & \\
\bottomrule
\end{tabular}
}
\end{table*}

\begin{table*}[!htp]\centering
\caption{The languages-abbreviation mapping of languages we used in low-resource tuning. }\label{tab: }
\resizebox{0.6\linewidth}{!}{
\begin{tabular}{lrrrrrrr}\toprule
Language &Serbian &Guarani &Hausa &Punjabi &Odia &Erzya \\
\midrule
Abbreviation &sr &gn &ha &pa &or &myv \\
\bottomrule
\end{tabular}
}
\end{table*}
\end{document}